\documentclass{article}

\usepackage{amssymb}
\usepackage{graphicx}
\usepackage{fullpage}
\usepackage{subfigure}
\usepackage{url}
\usepackage{algorithm}
\usepackage{algorithmic}

\renewcommand{\algorithmicrequire}{\textbf{Input:}}
\renewcommand{\algorithmicensure}{\textbf{Output:}}

\title{Embedding Graphs into the Extended Grid}
\author{Michael Coury 
\thanks{School of Computing Science, Simon Fraser University, Burnaby, BC, Canada, V5A 1S6 (mcoury@cs.sfu.ca)} 
} 

\begin{document}
\maketitle
\abstract{
Let $G=(V,E)$ be an arbitrary undirected source graph to be embedded in a target graph $EM$, the extended grid with vertices on integer grid points and edges to nearest and next-nearest neighbours.  We present an algorithm showing how to embed $G$ into $EM$ in both time and space $O(|V|^2)$ using the new notions of \textit{islands} and \textit{bridges}.  An island is a connected subgraph in the target graph which is mapped from exactly one vertex in the source graph while a bridge is an edge between two islands which is mapped from exactly one edge in the source graph.  This work is motivated by real industrial applications in the field of quantum computing and a need to efficiently embed source graphs in the extended grid.
}

\section{Introduction}
In this paper, we describe a method for embedding any undirected source graph into the extended grid.  We also introduce the concept of islands and bridges.  This embedding problem is of interest theoretically, and it has real industrial applications in the field of quantum computing that motivate this research.  We introduce a constructive algorithm to embed complete graphs in $O(n^2)$, thereby providing an upper bound.

An adiabatic quantum computer, such as one based on the system described by Amin, et alia \cite{misc:amin:2006}, can be considered to be a graph.  This allows for the computer to be programmed by formulating a given problem as a graph theoretic problem (e.g. cellular base station placement is formulated as maximum independent set) and then embedding the problem graph onto the graph representation of the quantum computer.  Therefore, the problem of embedding a source graph (representing the problem) into a target graph (representing the quantum computer architecture) is both interesting and important.

In collaboration with David Grant and William Macready, we developed the idea of a connected subgraph in the target graph to represent a single vertex in the source graph as well as a randomized $O(n^k), k > 2$, and then a $2n \times 2n$, algorithm for embedding.  Garey, Johnson, and So \cite{art:garey:1976} consider connected subgraphs of vertices, called nets, in the context of circuit testing; that is, looking for short circuits.  Opatrny and Sotteau \cite{art:opatrny:2000} and Lin, et alia \cite{art:sudborough:2003}, describe, in the contexts of VLSI and parallel computing, embedding complete binary trees into $EM$ using edge subdivision with vertex congestion of one, while Sang and Sudborough \cite{art:sang:1990} use vertex congestion and contraction to embed larger meshes into smaller ones.  Fraysseix, et alia \cite{art:fraysseix:1990} and Schnyder \cite{proc:schnyder:1990} describe $O(n^2)$ algorithms to obtain F\'{a}ry embeddings.

\subsection{Definitions}
We define the extended grid ($EM[m,n]$) to be an $m$ row by $n$ column lattice of grid points where every grid point has a potential edge to its immediate, or nearest, neighbours as well as to its next-nearest neighbours.  We will abbreviate $EM[m,n]$ as $EM$.  More specifically, $EM$ consists of the set of vertices given by the grid coordinates $\{ (x,y) | 1 \leq x \leq m, 1 \leq y \leq n \}$, and for a vertex $v = (x,y)$, the edges incident on $v$ are $\{ (u, v) | u \in (x, y \pm 1), (x \pm 1, y), (x \pm 1, y \pm 1) \}$.

Given an arbitrary source graph $G=(V,E)$, the problem is to embed this graph as a subgraph of $EM$, $\hat{G}=(\hat{V},\hat{E},\hat{C})$, where each $v \in V$ maps to a nonempty set $\vec{v} \subseteq \hat{V}$, each edge $(u,v) \in E$ maps to an edge $(u',v') \in \hat{E}$ with $u',v' \in \hat{V}, u' \in \vec{u}, v' \in \vec{v}$, and $\vec{v}$ is a connected subgraph whose edges are in $\hat{C}$.  That is, each vertex in the source graph maps to a set of vertices in the target graph that form a connected subgraph.  We call such a connected subgraph an island.  The edges in the island are in the edge set $\hat{C}$.  Each edge in the source graph also maps to an edge, which we will refer to as a \textit{bridge}, in the target graph's edge set $\hat{E}$ between two appropriate islands.

Put another way, let us define an island $I_v$ to be a connected subgraph of the embedding $\hat{G}$ where the vertices of $I_v$ collectively represent the vertex $v \in G$; i.e, the vertices of $I_v$ are $\vec{v}$.  Thus, we have a function $\eta_V : V \rightarrow \hat{V} \times \hat{V} \times ... \times \hat{V}$ that maps vertices to islands.  We then define a bridge to be an edge $(I_{u,s}, I_{v,t})$ connecting two islands $I_u$ and $I_v$, given by the bijective function $\eta_E : E \rightarrow \hat{E}$ that takes edges to bridges.

Given an input graph $G=(V,E)$, let us define an embedding of that graph to be a new graph $\hat{G}=(\hat{V},\hat{E},\hat{C})$. Also, let $l$ be a labelling function that maps a vertex of an island in the embedding to vertices in the input graph, $l : \hat{V} \rightarrow V$.  We can then say an island is
\[
\hat{V}_v = \{ v' | l(v') = v \} 
\]
and an edge set is
\[
\hat{C}_v = \forall u' \in \hat{V}_v\ \exists v' \in \hat{V}_v, \{ (u',v') | l(u')=l(v')=v \}.
\]  
From these definitions we get 
\begin{eqnarray*}
\hat{V} = \bigcup_{v\in V} \{ \hat{V}_v \} &,& \hat{C} = \bigcup_{v\in V} \{ \hat{C}_v \}.
\end{eqnarray*}
Furthermore, we define 
\[
\hat{E}_{uv} = \{ (u',v') | l(u') = u, l(v') = v \}
\]
and we get 
\[
\hat{E} = \bigcup_{(u,v)\in E} \{ \hat{E}_{uv}\}.
\]
Finally, we say $I_v = (\hat{V}_v, \hat{E}_v)$.

Thus, we create an embedding $\hat{G}$ of the input graph $G$ such that each vertex $v \in V$ corresponds to an island, $\hat{V}_v \in \hat{V}$, where $|\hat{V}_v| \geq 1$, with a set of edges $\hat{C}_v$ such that the graph $\hat{G}_v=(\hat{V}_v,\hat{C}_v)$ is connected.  Furthermore, for every edge $(u,v) \in E$ there exists an edge $(u',v') \in \hat{E}$ such that $l(u') = u, l(v') = v$. 

\section{Algorithm}\label{sec:upperBounds}
We now provide an $O(n^2)$ algorithm which will show how to embed any complete graph into the extended grid $EM[n-1,n]$ using islands and bridges, where $n = |V|$.  The embedding technique we describe will be call a braiding since we braid the layout of the islands together in the embedding.  One way to think of it is as a series of swaps, whereby we maximize the number of swaps, and the number of new edges introduced, at each row transition.

The algorithm is as follows.  For each odd-labelled vertex, we grow that island toward the right until it hits the boundary; at this point, the island reflects after a delay.  Even-labelled vertices are treated similarly except the initial growth is towards the left.  (Note that this orientation is arbitrarily chosen.)

We iterate through the rows, starting from the first.  Given a numbered ordering $n_V = \{ 1,2,...,n \}$ of $V$, we layout all the vertices in ascending order on the first row.  We proceed to layout the remaining rows as follows:

Let $c_i(v)$ denote the column in row $i$ for vertex $v$.  For each $v \in V$ and for each row, if $n_V(v)$ is odd, then $c_i(v) = c_{i-1}(v) - 1$ while $c_i(v) > 0$.  If $c_{i-1}(v) = 1$, then let $c_i(v) = 1$ and let $j=i$.  For all subsequent $i > j$, $c_i(v) = c_{i-1}(v) + 1$.  If $n_V(v)$ is even, the layout is reversed.

Thus,
\[
	c_i(v) = \left\{
		\begin{array}{ll}
		n_V(v) + h - \lfloor \frac{n_V(v)+h-1}{n} \rfloor (2r_o + 1) & \mbox{if $n_V(v)$ is odd}\\
		n_V(v) - h + \lceil \frac{-(n_V(v)-h-1)}{n} \rceil (2r_e + 1) & \mbox{if $n_V(v)$ is even}\\
		\end{array}
	\right.
\]
where
\begin{eqnarray*}
h   &=& i - 1, \\
r_o &=& h - (n - n_V(v)) - 1, \\
r_e &=& h - n_V(v).
\end{eqnarray*}

\begin{algorithm}
\caption{Braiding algorithm}\label{alg:braiding}

\begin{algorithmic}
\REQUIRE{A graph $G=(V,E)$}
\ENSURE{An embedding $\hat{G}=(\hat{V},\hat{E},\hat{C})$}
\medskip
    \STATE $m = |V| - 1$
    \STATE $n = |V|$
    
     \FOR{$nv=1$ to $n$}
         \STATE $i = 1$
         \STATE $j = nv$
         \IF{isOdd($nv$)}
             \STATE $left =$ \textbf{true}
         \ELSE
             \STATE $left =$ \textbf{false}
         \ENDIF
         \WHILE{$i \leq m$}
             \STATE $\hat{V}(i,j) = nv$
             \STATE $jlast = j$
             \STATE $i = i+1$
             \IF{left}
                 \STATE $j = j+1$
             \ELSE
                 \STATE $j = j-1$
             \ENDIF
             \IF{$j > n \wedge left$}
                 \STATE $j = n$
                 \STATE $left =$ \textbf{false}
             \ELSIF{$j < 1 \wedge \neg left$}
                 \STATE $j = 1$
                 \STATE $left =$ \textbf{true}
             \ENDIF

             \IF{$i \leq m$}
                 \STATE $\hat{C}($toIndex$(i-1,jlast),$toIndex$(i,j)) = 1$
             \ENDIF
         \ENDWHILE
     \ENDFOR

     \FOR{$i=1$ to $m$}
         \STATE $uv = \hat{V}(i,\{ 1...n \})$
         \FOR{$j=1$ to $|uv|-1$}
             \IF{$E(uv(j),uv(j+1)) \not= 0$}
                 \STATE $\hat{E}($toIndex$(i,j),$toIndex$(i,j+1)) = 1$
                 \STATE $\hat{E}($toIndex$(i,j+1),$toIndex$(i,j)) = 1$
                 \STATE $E(uv(j),uv(j+1)) = 0$
                 \STATE $E(uv(j+1),uv(j)) = 0$
             \ENDIF
         \ENDFOR
     \ENDFOR
\end{algorithmic}
\end{algorithm}

\begin{algorithm}
\caption{toIndex function}\label{alg:toIndex}

\begin{algorithmic}
\REQUIRE{the row $r$ and the column $c$}
\ENSURE{the index, $i$}
\STATE $i = (r-1) \cdot n + c$
\end{algorithmic}
\end{algorithm}

\begin{figure}[ht]
\centering
\subfigure{\includegraphics[width=0.45\textwidth]{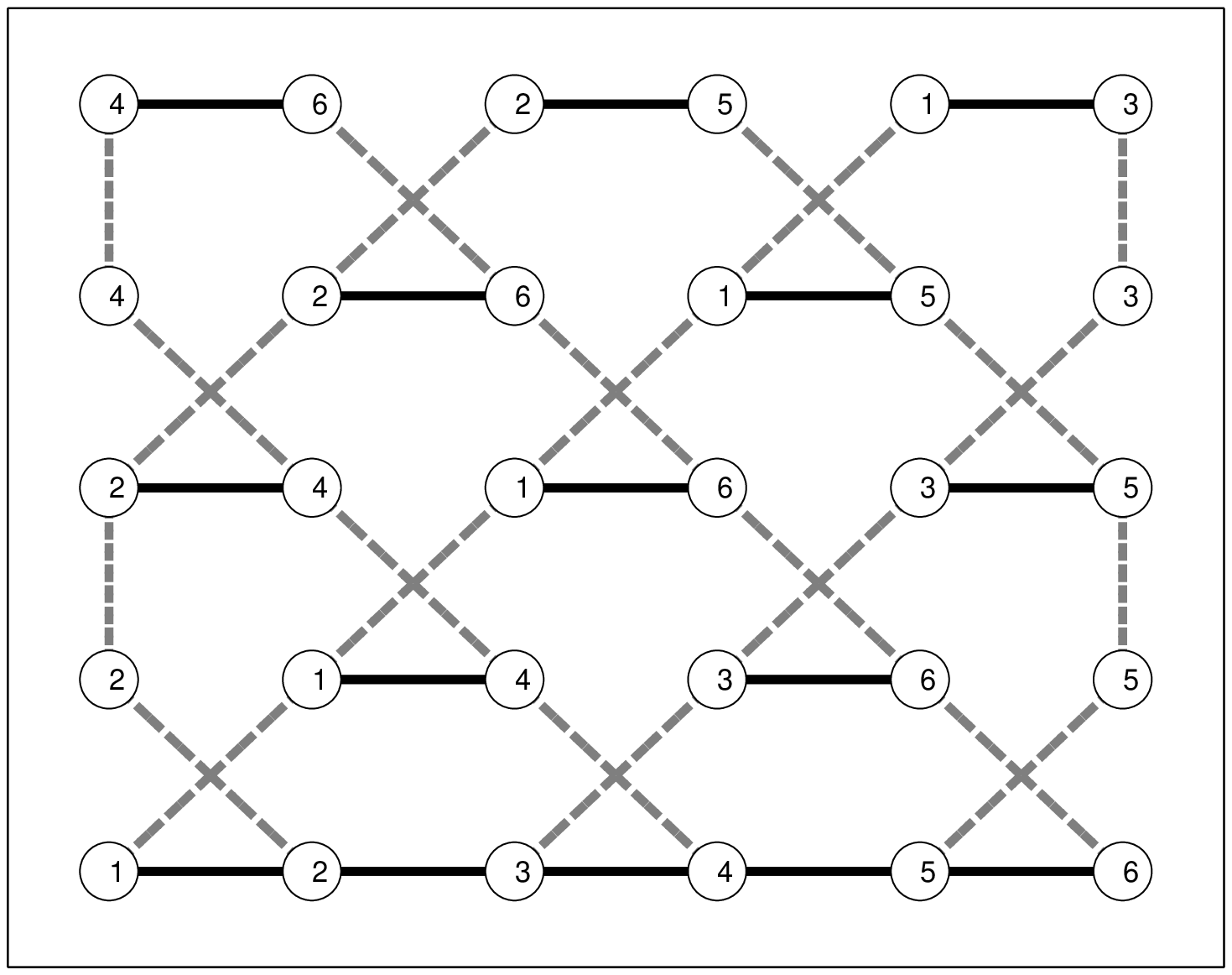}}
\subfigure{\includegraphics[width=0.45\textwidth]{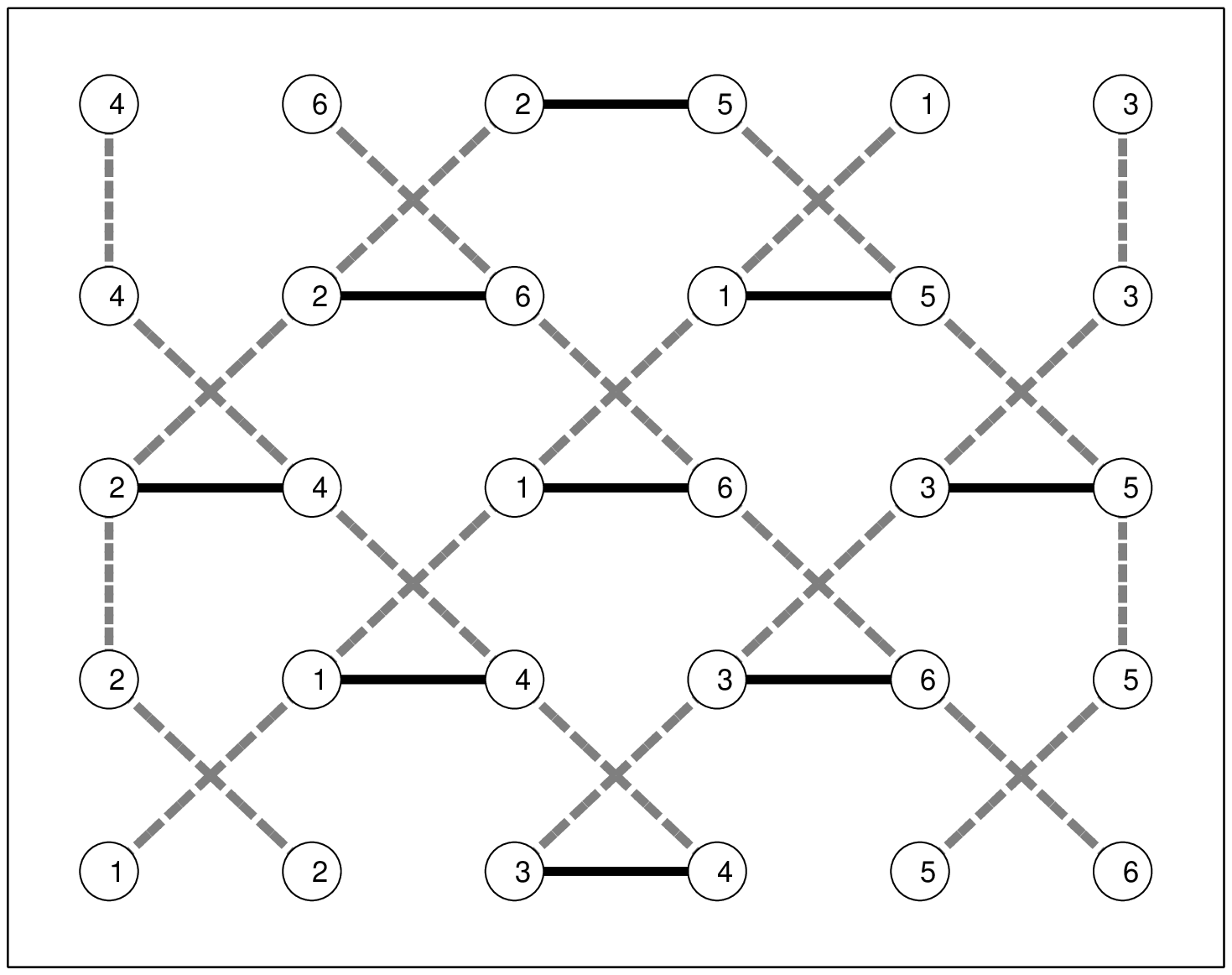}}
\caption{Braided embedding of (a) $K_6$, and (b) $K_{3,3}$\label{fig:braiding}}
\end{figure}

The algorithm, shown in Algorithm \ref{alg:braiding}, is quite simple and creates the braided embedding, as shown for $K_6$ and $K_{3,3}$ in Figure \ref{fig:braiding}, and is scalable to any number of vertices.  It is apparent that this is sufficiently scalable to embed any complete graph and therefore any graph.

\section{Acknowledgements}
The author would like to thank David Grant and Bill Macready for enlightening discussions.  We would also like to thank Joan and Anthony Geramita, and John Coury for helpful comments.

\bibliographystyle{elsart-num}

\end{document}